\documentclass[aps,pra,twocolumn,superscriptaddress]{revtex4}

\usepackage[latin1]{inputenc}
\usepackage[T1]{fontenc}
\usepackage[english]{babel}
\usepackage[dvips]{graphicx}
\usepackage{amsmath}
\usepackage{enumerate,amsthm,amsmath,amssymb,color,graphicx,bbm,float}
\usepackage{natbib}

\newtheorem{theo}{Theorem} 
\newtheorem{lemma}[theo]{Lemma}

\newcommand\bbone{\ensuremath{\mathbbm{1}}}

\begin{document}

\title{Multidimensional reconciliation for a continuous-variable
  quantum key distribution}

\author{Anthony Leverrier} \affiliation{Institut T\'el\'ecom /
  T\'el\'ecom ParisTech (Ecole Nationale Supérieure des
  T\'el\'ecommunications), CNRS LTCI, 46, rue Barrault, 75634 Paris
  Cedex 13, France} \author{Romain All\'eaume} \affiliation{Institut
  T\'el\'ecom / T\'el\'ecom ParisTech (Ecole Nationale Supérieure des
  T\'el\'ecommunications), CNRS LTCI, 46, rue Barrault, 75634 Paris
  Cedex 13, France} \author{Joseph Boutros} \affiliation{Texas $A\&M$
  University at Qatar, Doha, Qatar} \author{Gilles Z\'emor}
\affiliation{Institut de Math\'ematiques de Bordeaux, Universit\'e de
  Bordeaux 1, Bordeaux, France} \author{Philippe Grangier}
\affiliation{Laboratoire Charles Fabry, Institut d'Optique, CNRS,
  Universit\'e Paris-Sud, Campus Polytechnique, RD 128, 91127
  Palaiseau Cedex, France}

\date{\today}

\begin{abstract}
  We propose a method for extracting an errorless secret key in a
  continuous-variable quantum key distribution protocol, which is
  based on Gaussian modulation of coherent states and homodyne
  detection. The crucial feature is an eight-dimensional
  reconciliation method, based on the algebraic properties of
  octonions.  Since the protocol does not use any postselection, it
  can be proven secure against arbitrary collective attacks, by using
  well-established theorems on the optimality of Gaussian attacks. By
  using this new coding scheme with an appropriate signal to noise
  ratio, the distance for secure continuous-variable quantum key
  distribution can be significantly extended.
\end{abstract}


\maketitle

\section{Introduction}

A major practical application of quantum information science is
quantum key distribution (QKD) \cite{gisin:rmp}, which allows two
distant parties to communicate with absolute privacy, even in the
presence of an eavesdropper. Most QKD protocols encode information on
discrete variables such as the phase or the polarization of single
photons and are currently facing technological challenges, especially
the limited performances of photodetectors in terms of speed and
efficiency in the single photon regime. A way to relieve this
constraint is to encode information on continuous variables such as
the quadratures of coherent states \cite{grosshans:nature} which are
easily generated and measured with remarkable precision by standard
optical telecommunication components. In such a protocol, Alice draws
two random values $X_A,P_A $ with a Gaussian distribution
$\mathcal{N}(0,V_A)$ and sends a coherent state centered on
$(X_A,P_A)$ to Bob. Bob then randomly chooses one of the two
quadratures and measures it with a homodyne detection. After the
measurement, he informs Alice of his choice of quadrature.  Alice and
Bob then share correlated continuous variables from which a secret key
can in principle be extracted, provided that the correlation between
the shared data is high enough. This condition is the equivalent of
the maximal error rate allowed for the BB84 protocol for example
\cite{bb84}.

Currently, the main bottleneck of continuous-variable protocols lies
in the classical post-processing of information, more precisely in the
{\em reconciliation} step which is concerned with extracting all the
available information from the correlated random variables shared by
the legitimate parties at the end of the quantum part of the
protocol. This classical step must not be underestimated since an
imperfect reconciliation limits both the rate and the range of the
protocol.

Two different approaches have been used so far to extract binary
information from Gaussian variables. {\em Slice reconciliation}
\cite{VanAssche2004,Bloch2006} consists in quantizing continuous
variables and then correcting errors on these discrete variables. It
allows in principle to transmit more than 1 bit per pulse, and to
extract all the information available, but only if the quantization
takes place in $\mathbb{R}^d$ with $d\gg1$, which results in an
unacceptable increase of complexity in practice. Therefore the present
protocols use $d=1$, resulting in finite efficiency, which limits the
range to about 30 km. The second approach uses the sign of the
continuous variable to encode a bit, and it has the advantage of
simplicity. It can also be efficient, at least in the case where the
signal to noise ratio is low enough, so that less than $1$ bit per
pulse can be expected. But since the Gaussian distribution is centered
around $0$ and most of the data have a small absolute value, it
becomes difficult to discriminate the sign when the noise is
important. As a consequence, it has been proposed to use {\em
  post-selection} \cite{ralph, silberhorn2002,
  R04,namiki-2004,2005no-,Heid2007} to get rid of the "low amplitude"
data, and keep only the more meaningful "large amplitude"
data. However, this approach has a major drawback: since the optimal
attack against such a post-selected protocol is unknown, the secret
rate can be calculated only for certain types of "restricted" attacks
\cite{silberhorn2002, Heid2007}. So the security is significantly
weaker than the initial "non post-selected" Gaussian-modulated
protocol, where one can use the optimality of Gaussian attacks
\cite{garcia-patron:prl, navascues:prl06} in order to prove that the
protocol is secure against arbitrary general collective attacks.

\begin{figure}
  \begin{center}
    \includegraphics[width=0.45\textwidth]{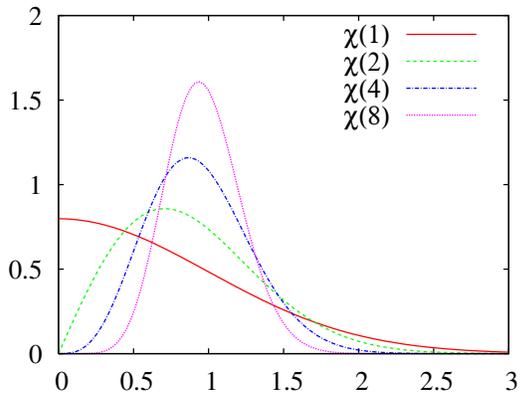}
    \caption{\small{(Color online) Probability distributions $\chi(1),
        \chi(2), \chi(4), \chi(8)$ of the radius of a Gaussian vector
        of dimension 1, 2, 4 and 8. When the dimension goes to
        infinity, the distribution gets closer to a Dirac
        distribution.} }
    \label{Chi distribution}
  \end{center}
\end{figure}

\begin{figure}
  \begin{center}
    \includegraphics[width=0.45\textwidth]{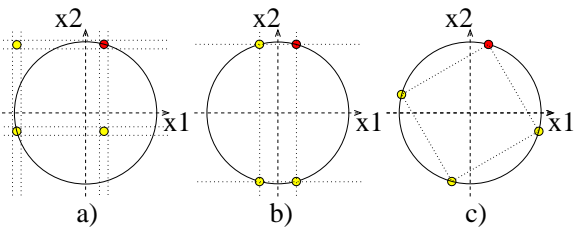}
    \caption{\small{(Color online) Consider two successive states
        $X_1, X_2$ sent by Alice: the states really sent correspond to
        $X_1>0, X_2>0$. Figures a), b) and c) show the four possible
        states Bob needs to discriminate after Alice has sent him some
        side information over the classical authenticated channel. a)
        corresponds to slice reconciliation
        \cite{VanAssche2004,Bloch2006}: the four states are well
        separated but the Gaussian symmetry is broken, b) corresponds
        to the case where the information is encoded on the sign of
        the Gaussian value \cite{silberhorn2002}}: the symmetry of the
      problem is preserved but some states are very close and thus
      difficult to discriminate, c) corresponds to the approach
      presented in this paper where the states are well separated and
      the symmetry is preserved. }
    \label{Chi}
  \end{center}
\end{figure}

Here we are interested in the problem of extending continuous-variable
QKD over longer distances without post-selection, but with proven
security. The main idea is as follows : whereas Gaussian random values
are centered around $0$, this is not the case for the norm of a
Gaussian random vector. Such a vector lies indeed on a shell which
gets thinner as the dimension of the space increases (see Fig.
\ref{Chi distribution}). Thus, if one performs a clever rotation (see
Fig. \ref{Chi}) before encoding the key in the sign of the
coordinates, one automatically gets rid of the small absolute value
coordinates without post-selection. Whereas this effect gets stronger
and stronger for large dimensions, we will show that we are
intrinsically limited to performing such rotations in
$\mathbb{R}^8$. As we will show below, this is related to the
algebraic structure of octonions. For our purpose, working in
$\mathbb{R}^8$ is already a significant improvement since it allows to
exchange secure secret keys over more than 50 km, without
post-selection, and with a reasonable complexity for the
reconciliation protocol.

The paper is organized as follows: Section \ref{security} presents the
link between the reconciliation and the security of the protocol,
Sec. \ref{binary} describes the reconciliation in the case of discrete
variables QKD protocols, Sec. \ref{gaussian} shows how to generalize
this approach to Gaussian variables protocols, and
Sec. \ref{rotations} presents a realistic reconciliation protocol for
continuous-variable QKD, whose performance is analyzed in
Sec. \ref{qkd}.

\section{Reconciliation and security}
\label{security}

Let $x$ and $y$ be the classical random variables associated with the
measured quantities of the legitimate parties Alice and Bob, and let
$E$ be the quantum state in possession of the eavesdropper. It has
been shown \cite{garcia-patron:prl, navascues:prl06} that the
theoretical secret key rate $K$ obtained using one-way reconciliation
is bounded from below by
$$K\geq I(x:y)-S(x:E)\equiv K_{\text{th}}.$$
Here $I(x:y)$ and $S(x:E)$ refer, respectively, to the Shannon mutual
information \cite{cover} between classical random values $x$ and $y$
and to the quantum mutual information \cite{Nielsen} between $x$ and
the quantum state $E$. Recall that $S(x:E)$ can also be seen as the
Holevo quantity associated to the quantum measurements performed by
Eve. The above bound corresponds to the case where Alice and Bob are
"classical" whereas Eve is "quantum", which means that Eve is allowed
to use a quantum memory and a quantum computer to perform her
attack. This secret key rate is valid for one-way reconciliation: the
classical communication between Alice and Bob is therefore restricted
to be unidirectional, and not interactive. For the protocol described
above, the quantum mutual information between Bob and Eve is smaller
than between Alice and Eve. As a consequence, one will use reverse
reconciliation \cite{grosshans:nature}: the final key is extracted
from Bob's data, and Bob sends extra information to Alice on the
authenticated classical channel to help her correct her "errors".  The
secret key rate $K_{\text{th}}$ is secure against collective
attacks. Note that it is conjectured that, as it is the case for
discrete variables protocols \cite{R05}, coherent attacks are not more
powerful than collective attacks \cite{garcia-patron:prl,
  navascues:prl06, renner2007}, which would imply that $K_{\text{th}}$
is the secure key rate against the most general attacks allowed by
quantum mechanics.

An important property of the continuous-variable QKD is that for a
reasonably low excess noise (which is the noise not directly caused by
the losses), $K_{\text{th}}$ remains strictly positive for any value
of the transmission meaning that there is not any theoretical
limitation to the range of this protocol. However $K_{\text{th}}$ is
relevant only in the case where one has access to a perfect
reconciliation scheme, allowing Alice and Bob to extract all the
information available in their correlated data. How should
$K_{\text{th}}$ be modified in the case of a real-world imperfect
reconciliation scheme?  In order to extract a secret from their data,
Alice and Bob have access to a classical authenticated channel and
have agreed on a particular code $\mathcal{C}_N$ whose size $N$ is
such that $\log_2(N) \leq I(x;y)$. The principle of the reconciliation
protocol is the following: Alice chooses randomly an element $U \in
\mathcal{C}_N$ and sends some information $\alpha$ to Bob who should
be able to efficiently recover $U$ from the knowledge of $y$ and
$\alpha$, i.e., $H(U|y,\alpha)=0$, the conditional entropy of $U$
given $y$ and $\alpha$ is null, or equivalently $I(U:y,\alpha)=H(U)$.
In this case, Alice and Bob have extracted a common string $U$ from
their data, which they will be able to turn into a secret key thanks
to privacy amplification, but they have also given the extra
information $\alpha$ to the eavesdropper. As a consequence, the
effective key rate after the reconciliation becomes:
$$K\geq H(U)-S(U:E,\alpha)\equiv K_{\text{real}}.$$
Unfortunately, one always has $K_{\text{real}} < K_{\text{th}}$ and
$K_{\text{real}}$ reaches $0$ for a finite channel transmission. In
other words, the range of the protocol is limited because of the
imperfect reconciliation. It should be noted that this is one of the
main differences with discrete variables protocols which are limited
by technology, and more particularly by the dark counts of the
photodetectors.  A real difficulty lies in the estimation of
$S(U:E,\alpha)$. One specificity of QKD is that it allows Alice and
Bob to estimate an upper bound of $S(x:E)$ by comparing a subset of
their data. However it is generally impossible to deduce
$S(U:E,\alpha)$ from it. One exception is when $U$ and $\alpha$ are
independent, in which case the following lemma applies.

\begin{lemma}
  \label{independence}
  Let $A$ and $B$ be two classical random values, let $E$ be a random
  quantum state. If $A$ and $B$ are independent, then $S(A:E,B)\leq
  S(A,B:E)$.
\end{lemma}

\begin{proof}
  The chain rule for mutual quantum information reads:
 $$S(A,B:E) =  S(B:E)+S(A:E|B) \geq  S(A:E|B) $$ 
 where the inequality results from the non-negativity of mutual
 quantum information. Then, by definition of conditional mutual
 information,
 \begin{eqnarray*}
   S(A:E|B)&=& S(A|B)-S(A|E,B) = S(A)-S(A|E,B)\\
   &=& S(A:E,B)
 \end{eqnarray*}
 where the second equality follows from independence of $A$ and $B$.
\end{proof}

In the reconciliation protocol, $U$ is chosen randomly by Alice,
independently of $x$, meaning that $S(x,U:E)=S(x:E)$. Then, since
$\alpha$ is a function of $x$ and $U$, the data-processing inequality
gives $S(U,\alpha:E)\leq S(x:E)$. In addition, in the case where
$\alpha$ is independent of $U$, lemma (\ref{independence}) gives:
$S(U:E,\alpha)\leq S(x:E).$

If one defines the efficiency of reconciliation $\beta =
\frac{H(U)}{I(x:y)}$, one obtains finally
$$K_{\text{real}} \geq \beta I(x:y)-S(x:E),$$
which is the usual expression of the secret key rate taking into
account the imperfect reconciliation protocol.

\section{Reconciliation of binary variables}
\label{binary}

Reconciliation is a means for Alice and Bob to extract available
common information from their correlated data. In the case when the
data consists of binary strings, it is very similar to the problem of
channel coding where the goal is for Alice to send information to Bob
through a noisy channel. Channel coding is solved by appropriately
choosing subsets of binary strings: codes. When Alice restricts her
messages to code words, Bob can recover them with high probability if
the code size is not too large, given the channel noise. More
precisely, Shannon's theorem \cite{Sha48} states that the size of the
code $|\mathcal{C}|$ is bounded by the mutual information between
Alice and Bob: $\log_2(|\mathcal{C}|)\leq I(x:y)$. The problem of
channel coding has been extensively studied during the past 60 years,
but only recently were discovered codes almost achieving Shannon's
limit while being efficiently decoded thanks to iterative algorithms:
turbocodes \cite{berrou} and Low Density Parity Check (LDPC) codes
\cite{Richardson2001a}.

The main difference between reconciliation and channel coding is that
in the case of reconciliation, Alice does not choose what she sends
and thus cannot restrict her messages to code words of a given
code. However, if one wants to take advantage of the code formalism,
knowing what she sent, Alice can describe to Bob a code for which her
word is a code word. Thus if Bob can guess what codeword Alice sent,
they will effectively share a common sequence of bits. This is the
method used for discrete QKD protocols. Indeed, given a linear code
$\mathcal{C}$ and its parity check matrix $H$, the group
$\mathbb{F}_2^n= \{0,1 \}^n$ of possible states sent by Alice can be
seen as the product of code words and syndromes: if Alice sends $x$ to
Bob, she can tell him the syndrome of $x$ which is $H \cdot x$ thus
defining a coset code containing $x$. This coset code is the ensemble:
$\{y \in \mathbb{F}_2^n |H \cdot y = x\}$.  An equivalent solution is
for Alice to randomly choose a code word $U$ from a given code and to
send $U \oplus x = \alpha$ to Bob where $\oplus$ represents the
addition in the group $\mathbb{F}_2^n$. Bob then computes $y \oplus
\alpha$ which allows him to retrieve $U$ if the code is well adapted
to the channel between Alice and Bob. This coset coding scheme was
initially suggested by Wyner \cite{Wyner}.

In a way, the side information (information sent by Alice over the
classical authenticated channel) corresponds to a change of
coordinates allowing one to transform the initial reconciliation
problem into the well-known problem of channel coding.

Two properties are essential for this approach to work: first, the
probability distribution of the states sent by Alice is uniform over
$\mathbb{F}_2^n$; second, the total space is a partition of the cosets
of a linear code. Thus, any word can be seen as a unique codeword for
a unique coset code and telling which coset code contains the word
gives zero information about the codeword. The question is then
whether or not it is possible to generalize this approach to
continuous variables.

\section{Reconciliation of Gaussian variables}
\label{gaussian}

\subsection{Gaussian modulation}

One of the main differences between discrete and continuous QKD
protocols is the probability distribution of Alice's variables: the
uniform distribution on $\mathbb{F}_2^n$ is changed into a nonuniform
Gaussian distribution on $\mathbb{R}^n$. This is rather unfortunate
since the uniformity of the distribution on $\mathbb{F}_2^n$ is an
essential assumption in order to prove that the side information
(e.g., the syndrome) Alice sends to Bob on the public channel does not
give any relevant information to Eve about the code word chosen by
Alice.  An interesting property of the Gaussian distribution
$\mathcal{N}(0, \bbone_n)$ on $\mathbb{R}^n$ whose covariance matrix
is the identity is that it has a spherical symmetry in
$\mathbb{R}^n$. In other words, if the vector $x$ follows such a
distribution, then the normalized random vector $\frac{x}{|x|}$ has a
uniform distribution on the unit sphere $\mathcal{S}^{n-1}$ of
$\mathbb{R}^n$. Thus, spherical codes, codes for which all codewords
lie on a sphere centered on $0$, can play the same role for
continuous-variable protocols as binary codes for discrete
protocols. Some very good codes are known for binary channels: LDPC
codes and turbocodes both almost achieve the Shannon limit and can be
efficiently decoded thanks to iterative decoding algorithms. Are there
codes with similar qualities among the spherical codes? The answer is
almost. There is indeed a canonical way to convert binary codes into
binary spherical codes and this can be achieved thanks to the
following mapping of $\mathbb{F}_2^n$ onto an isomorphic image in the
$n$-dimensional sphere:
$$  \mathbb{F}_2^n  \rightarrow  \mathcal{S}^{n-1} \subset
\mathbb{R}^{n}, (b_1, \dots, b_n) \mapsto \left(
  \frac{(-1)^{b_1}}{\sqrt{n}}, \dots,\frac{(-1)^{b_n}}{\sqrt{n}}
\right).$$ Then, as LDPC codes and turbocodes can both be optimized
for binary symmetric channels, they can also be optimized for a binary
phase shift keying (BPSK) modulation, where the bit $0$ $(1)$ is
encoded into the amplitude $+A$ $(-A)$, and where the channel noise is
considered to be additive white Gaussian noise (AWGN). Thus, one has
access to a family of very good codes (in the sense that they are very
close to the Shannon limit) for which very efficient iterative
decoding algorithms are available. It is important to note that there
are actually two different Shannon limits considered here depending on
the modulation, BPSK or Gaussian modulation, but these limits become
asymptotically close when the signal-to-noise ratio (SNR) is
small. Thus, at low SNR, a binary code optimized for a BPSK modulation
can almost achieve the Shannon limit for a Gaussian modulation.

A remark is in order~: the use of binary codes as described above
limits the rate of the code to less than $1$ bit per channel use,
whereas one of the interests of a Gaussian modulation is precisely to
get rid of this limit. Actually, one could use nonbinary spherical
codes, but their decoding is more complicated and thus slows down the
reconciliation protocol. In addition, this is not really needed, since
in the high loss scenario which interests us most here, the secret key
rate is always much less than 1 bit per channel use. Consequently the
use of binary codes turns into an advantage, since they can be decoded
very efficiently.  In the low-loss case however, that is for short
distances, one can hope to distill more than $1$ bit per channel use,
and the "usual" approach \cite{lodewyck:pra07} will be more suitable
than the one described in the present article (see also discussion in
Sec. \ref{qkd}).

Now that we have a probabilistic space with a uniform probability
distribution and a family of codes for this space, we need to see if
the total space is a partition of a code and of its "generalized coset
codes". First, the canonical hypercube of $\mathbb{R}^{n}$ (which is
the image of $\mathbb{F}_2^n$ by the isomorphism defined above) is
described as a partition of a linear code and its cosets. The question
that remains to be solved is whether or not the unit sphere is a
partition of such hypercubes. Another way to see this problem is the
following: given a random point in $\mathcal{S}^{n-1}$, is there a
hypercube inscribed in the sphere for which this point is a
vertex. Surely there are such hypercubes, many in fact. Actually, the
manifold of these hypercubes is a $[(n-1)(n-2)/2]$-dimensional
manifold (this is the dimension of the subgroup of orthogonal group
$O_n$ that transports the canonical hypercube onto the ensemble of
hypercubes containing the point in question).

Yet another way to express the problem is the following: given two
points $x,y \in \mathcal{S}^{n-1}$, is it possible to find an
orthogonal transformation mapping $x$ to $y$? One can immediately
think of transformations such as the reflection across the mediator
hyperplane of $x$ and $y$. Unfortunately, such an orthogonal
transformation gives some information about $x$ and $y$ as soon as
$n>2$ (this is linked to the phenomenon of concentration of measure
for spheres in dimensions $n>2$), and therefore cannot be used by
Alice as legitimate side information, which should be independent from
the key in order to fulfill the hypothesis of Lemma
\ref{independence}.

A correct solution would then be to randomly choose an orthogonal
transformation with uniform probability in the ensemble of orthogonal
transformations mapping $x$ to $y$. This can be done in the following
way: one first draws a random orthogonal transformation mapping $x$ to
some random $x'$. Then one composes this transformation with the
reflection across the mediator hyperplane of $x'$ and $y$. Although
theoretically correct, this procedure is not doable in practice for
$n\gg 1$ since generating a random orthogonal transformation on
$\mathbb{R}^{n}$ is a computational demanding task requiring to draw
an $n \times n$ Gaussian random matrix and to calculate its QR
decomposition (i.e., its decomposition into an orthogonal and a
triangular matrix) which is an operation of complexity $O(n^3)$.

A practical solution involves the following: for each word $x \in
\mathcal{S}^{n-1}$ sent by Alice, for each code word $U \in
\mathcal{S}^{n-1}$ chosen by Alice (not necessarily a binary
codeword), there should exist an continuous application $M$ of the
variables $x$ and $U$ such that $M(x,U) \in O_n$ and $M(x,U) \cdot x =
U$. Then if Alice gives $M(x,U)$ to Bob, one has the continuous
equivalent of $U \oplus x$ in the discrete protocol. The following
theorem shows that the existence of such an application $M$ restricts
the possible values of $n$ to be $1$, $2$, $4$ or $8$.

\begin{theo}
  \label{Existence of M}
  If there exists a continuous application
$$ M : \mathcal{S}^{n-1} \times \mathcal{S}^{n-1}  \rightarrow  O_n, 
(x,y) \mapsto M(x,y) $$ such that $M(x,y) \cdot x = y$ for all $x, y
\in \mathcal{S}^{n-1}$, then $n =$ $1$, $2$, $4$ or $8$.
\end{theo}

The proof of this theorem uses a result from Adams \cite{adams}, which
quantifies the number of independent vector fields on the unit sphere
of $\mathbb{R}^n$:
\begin{theo}
  \label{Adams}
  Independent vector fields on $\mathcal{S}^{n-1}$ (J.F. Adams,
  1962). For $n=a \cdot 2^b$ with $a$ odd and $b=c+4d$, one defines
  $\rho_n=2^c+8d$. Then the maximal number of linearly independent
  vector fields on $\mathcal{S}^{n-1}$ is $\rho_n-1$.
\end{theo}
In particular, the only spheres for which there exist $(n-1)$
independent vector fields are the unit sphere of $\mathbb{R}$,
$\mathbb{R}^2$, $\mathbb{R}^4$ and $\mathbb{R}^8$, which can
respectively be seen as the units of the real numbers, the complex
numbers, the quaternions and the octonions.

\begin{proof} [Proof of Theorem \ref{Existence of M}.] The idea of the
  proof is to use the existence of such a continuous function $M$ to
  exhibit a family of $(n-1)$ independent vector fields on $S^{n-1}$.

  Let $(e_1, e_2, \dots, e_n)$ be the canonical orthonormal basis of
  $\mathbb{R}^n$.  For $1\leq i\leq n$, let $u_i(x) = M(e_n,x) \cdot
  e_i$.  One has: $u_n(x)=x$ and
  \begin{eqnarray*}
    (u_i(x) | u_j(x))  & = & e_i^T M(e_n,x)^T M(e_n,x) e_j \\
    & = & \delta_{i,j} \hspace{1cm} \text{since $M(e_n,x) \in O_n$} \\
  \end{eqnarray*}

  Then, for $x \in \mathcal{S}^{n-1}$, $u_1(x), u_2(x), \dots,
  u_{n-1}(x)$ are $(n-1)$ independent vector fields on
  $\mathcal{S}^{n-1}$ and finally $n = 1, 2, 4$ or $8$.
\end{proof}

\section{Rotations on $\mathcal{S}^1$, $\mathcal{S}^3$ and
  $\mathcal{S}^7$}
\label{rotations}

Now that we have proved that such an application $M$ can only exist in
$\mathbb{R}$, $\mathbb{R}^2$, $\mathbb{R}^4$ and $\mathbb{R}^8$, we
need to answer three more questions: does it exist? Can Alice compute
it efficiently? Does it leak any information about the codeword to
Eve?  Note that the trivial case of $\mathbb{R}$ for which the unit
sphere is $\{-1,1\}$ corresponds to the method where one encodes a bit
in the sign of the Gaussian variable \cite{silberhorn2002}.

\subsection{Existence}

Let us start with the easiest case: $\mathbb{R}^2$. The existence of
such an application $M$ verifying $M(x,y) \cdot x = y$ for the unit
circle is obvious: it is simply the rotation centered in $O$ of angle
$\text{Arg}(y)-\text{Arg}(x)$ where Arg$(x)$ denotes the angle between
$x$ and the x-axis. An alternative way to see $M$ is $M(x,y)=yx^{-1}$
where $x$ and $y$ are identified with complex numbers of modulus $1$.
The same is true for dimensions $4$ and $8$ where $\mathcal{S}^3$ and
$\mathcal{S}^7$ can respectively be identified with the quaternion
units and the octonion units, and for which a valid division exists.

\subsection{Computation of $M(x,y)$}

For $n=2,4$ and $8$, there exists a (nonunique) family of $n$
orthogonal matrices $\mathcal{A}_n=(A_1, \dots, A_n)$ of
$\mathbb{R}^{n \times n}$ such that $A_1=\bbone_n$, and for $i,j>1$,
$\{A_i,A_j\}=-2 \delta_{i,j} \bbone_n$ where $\{A,B\}$ is the
anticommutator of $A$ and $B$. An example of these families is
explicitly given in the Appendix. The following lemma shows how to use
such a family to construct a continuous function $M$ with the
properties described above.
\begin{lemma}
  \label{definition of M}
  $ \displaystyle {M(x,y)=\sum_{i=1\dots n}\alpha_i(x,y) A_i}$ with
  $\alpha_i(x,y)=(A_i x |y)$ is a continuous map from
  $\mathcal{S}^{n-1} \times \mathcal{S}^{n-1}$ to $O(n)$ such that
  $M(x,y) x = y$.
\end{lemma}

\begin{proof}
  First, because of the anticommutation property, one can easily check
  that the family $(A_1x, A_2x, \dots, A_nx)$ is an orthonormal basis
  of $\mathbb{R}^n$ for any $x \in \mathcal{S}^{n-1}$.  Then, for any
  $x,y \in \mathcal{S}^{n-1}$, $(\alpha_1(x,y),\dots,\alpha_n(x,y))$
  are the coordinates of $y$ in the basis $(A_1x, A_2x, \dots,
  A_nx)$. This proves that $M(x,y) x = y$.  Finally, the orthogonality
  of $M(x,y)$ follows from some simple linear algebra.
\end{proof}

Then $\alpha=(\alpha_1,\dots,\alpha_n)$ is sufficient to describe
$M(x,y)$ and the computation of $\alpha_i$ can be done efficiently
since the matrices $A_i$ are just permutation matrices with a change
of sign for some coordinates.  In the QKD protocol, Alice chooses
randomly $u$ in a finite code and gives the value of $\alpha(x,u)$ to
Bob, who is then able to compute $M(x,u)y$ which is a noisy version of
$u$. One should note that the final noise is just a "rotated" version
of the noise Bob has on $x$: in particular, both noises are Gaussian
with the same variance.

\subsection{No leakage of information}

In order to prove that $\alpha = M(x,u)$ does not give any information
about $u$, one needs to show that $u$ and $\alpha$ are independent, in
other words that: $Pr(u=u_i|M(x,u)=\alpha)=Pr(u=u_i)=\frac{1}{N}$ if
one considers the spherical code $\mathcal{C}_N = \{u_1, \dots, u_N
\}$. This is true because $x$ and $u$ have uniform distributions (on
$\mathcal{S}^{n-1}$ and $\mathcal{C}_N$ respectively) and because the
function:
$$ f_u : \mathbb{R}^n  \rightarrow  \mathbb{R}^n,
x \mapsto f_u(x) = \alpha, \text{ with } \alpha_i=(u|A_i x)$$ has a
constant Jacobian equal to $1$ for each $u \in \mathcal{C}_N$. To see
this, one should note that the lines of the Jacobian matrix of $f_u$
are the $A_i^Tu$ which form an orthonormal basis of $\mathbb{R}^n$.

\section{Application to the continuous-variable QKD}
\label{qkd}

Now that we have explained how efficient reconciliation of correlated
Gaussian variables can be achieved with rotations in $\mathbb{R}^8$,
let us look at the implications for the continuous-variable QKD.

At the end of the quantum part of the continuous-variable QKD
protocol, Alice and Bob share correlated random values and their
correlation depends on the variance of the modulation of the coherent
states and on the properties of the quantum channel. The channel can
safely be assumed to be Gaussian since it corresponds to the case of
the optimal attack for Eve. This means that it can be entirely
characterized by its transmission and excess noise. Both these
parameters are accessible to Alice and Bob through an estimation step
prior to the reconciliation \cite{R05}. Once these parameters are
known, one can calculate the signal-to-noise ratio (SNR) of the
transmission, which is the ratio between the variance of the signal
(the variance of the Gaussian modulation of coherent states in our
case) and the variance of the noise (noise induced by losses as well
as excess noise).  The SNR quantifies the mutual information between
Alice and Bob when a Gaussian modulation is sent over a Gaussian
channel:
$$I(A:B)=\frac{1}{2} \log_2(1+\text{SNR}).$$
Note also that the efficiency of the reconciliation only depends on
the correlation between Alice's and Bob's data, that is, on the
SNR. Thus, for a given transmission and excess noise, the secret key
rate is a function of the SNR, which can be optimized by changing the
variance of the modulation of the coherent states.

It is not easy to know exactly how the efficiency of reconciliation
depends on the SNR. However, each reconciliation technique performs
better for a certain range of SNR: slice reconciliation is usually
used for a SNR around $3$ \cite{lodewyck:pra07} while rotations in
$\mathbb{R}^8$ are optimal for a low SNR, typically around $0.5$.

\begin{figure}
  \begin{center}
    \includegraphics[width=0.45\textwidth]{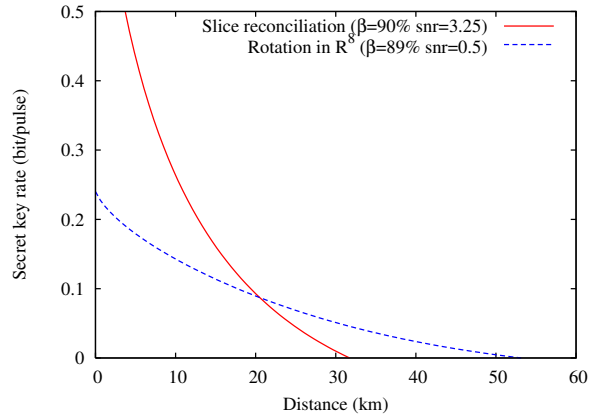}
    \caption{\small{(Color online) Performance of slice reconciliation
        vs rotation in $\mathbb{R}^8$. Experimental parameters: excess
        noise referred to the channel input $\xi=0.005$, efficiency of
        Bob's detector $\eta=0.606$ and electronic noise at Bob's side
        $V_{elec}=0.041$ \cite{lodewyck:pra07}. The reconciliation
        based on rotations in $\mathbb{R}^8$ uses a LDPC code of rate
        0.26 \cite{epfl}} }
    \label{Comparison}
  \end{center}
\end{figure}

Figure \ref{Comparison} shows the performance of rotations in
$\mathbb{R}^8$ compared to slice reconciliation for the experimental
parameters of the QKD system developed at Institut d'Optique. Both
approaches achieve comparable reconciliation efficiencies (around $90
\%$) but for different SNR. One can observe two distinct regimes: for
low loss, i.e., short distance, slice reconciliation is better but
only rotations in $\mathbb{R}^8$ allow QKD over longer distances (over
50 km with the current experimental parameters).

Concerning the complexity of the reconciliation, one should be aware
that almost all the computing time is devoted to decoding the
efficient binary codes, either LDPC codes or turbocodes. Compared to
this decoding, the rotation in $\mathbb{R}^8$ takes a negligible
amount of time. Thus, the complexity of the reconciliation presented
here is smaller than the one of slice reconciliation since the latter
uses several codes (one code per slice).

\section{Conclusion}
We presented a protocol for the reconciliation of correlated
Gaussian variables. Currently, the main bottleneck of
continuous-variable QKD lies in the impossibility for Alice and Bob to
extract efficiently all the information available, this difficulty
resulting in both a limited range and a limited rate for the key
distribution. The method described in this article is particularly
well adapted for low signal-to-noise ratios, which is the situation
encountered when one wants to perform QKD over long distances. By
taking into account the current experimental parameters of the QKD
link developed at the Institut d'Optique \cite{lodewyck:pra07}, one
shows that this new reconciliation allows QKD over more than 50 km.
Moreover, contrary to other protocols that have been proposed to
increase the range of continuous-variable QKD, this protocol does not
require any post-selection. Hence, the security proofs based on the
optimality of Gaussian attacks \cite{garcia-patron:prl,
  navascues:prl06} remain valid, meaning that the protocol is secure
against general collective attacks.

\begin{acknowledgments}
  We thank Thierry Debuisschert, Eleni Diamanti, Simon Fossier,
  Fr\'ed\'eric Grosshans and Rosa Tualle-Brouri for helpful
  discussions. We also acknowledge the support from the European Union
  under the SECOQC project(IST-2002-506813) and the French Agence
  Nationale de la Recherche under the PROSPIQ project
  (No. ANR-06-NANO-041-05.
\end{acknowledgments}

\appendix*

\section{Examples of families $\mathcal{A}_2$, $\mathcal{A}_4$ and
  $\mathcal{A}_8$}

\subsection{Notations} Let us introduce the following 4 $2 \times 2$ matrices:\\
$K_0=
\begin{pmatrix}
  1&0 \\
  0&1
\end{pmatrix}, K_1=
\begin{pmatrix}
  0&1 \\
  1&0
\end{pmatrix}, K_2=
\begin{pmatrix}
  0&-1 \\
  1&0
\end{pmatrix}$ and $K_3=
\begin{pmatrix}
  1&0 \\
  0&-1
\end{pmatrix}$ and the tensor product $K_{i_1,..,i_l}=K_{i_1} \otimes
.. \otimes K_{i_l}$.

\subsection{Examples}

Family $\mathcal{A}_2$: $\{K_{0},K_{2}\}$\\

$ A_1=
\begin{pmatrix}
  1&0 \\
  0&1
\end{pmatrix}, A_2=
\begin{pmatrix}
  0&-1 \\
  1&0
\end{pmatrix}$\\

Family $\mathcal{A}_4$: $\{K_{00},K_{32},K_{20},K_{12}\}$\\

$ A_1=
\begin{pmatrix}
  1&0&0&0 \\
  0&1&0&0 \\
  0&0&1&0 \\
  0&0&0&1 \\
\end{pmatrix}, A_2=
\begin{pmatrix}
  0&-1&0&0 \\
  1&0&0&0 \\
  0&0&0&1 \\
  0&0&-1&0 \\
\end{pmatrix},$

$A_3=
\begin{pmatrix}
  0&0&-1&0 \\
  0&0&0&-1 \\
  1&0&0&0 \\
  0&1&0&0 \\
\end{pmatrix}, A_4=
\begin{pmatrix}
  0&0&0&-1 \\
  0&0&1&0 \\
  0&-1&0&0 \\
  1&0&0&0 \\
\end{pmatrix}$\\

Family $\mathcal{A}_8$: $\{K_{000},K_{332},K_{320},K_{312},K_{200},K_{102},K_{123},K_{121}\}$\\

$ A_1=
\begin{pmatrix}
  1&0&0&0&0&0&0&0 \\
  0&1&0&0&0&0&0&0 \\
  0&0&1&0&0&0&0&0 \\
  0&0&0&1&0&0&0&0 \\
  0&0&0&0&1&0&0&0 \\
  0&0&0&0&0&1&0&0 \\
  0&0&0&0&0&0&1&0 \\
  0&0&0&0&0&0&0&1 \\
\end{pmatrix},$

$A_2=
\begin{pmatrix}
  0&-1&0&0&0&0&0&0 \\
  1&0&0&0&0&0&0&0 \\
  0&0&0&1&0&0&0&0 \\
  0&0&-1&0&0&0&0&0 \\
  0&0&0&0&0&1&0&0 \\
  0&0&0&0&-1&0&0&0 \\
  0&0&0&0&0&0&0&-1 \\
  0&0&0&0&0&0&1&0 \\
\end{pmatrix},$

$A_3=
\begin{pmatrix}
  0&0&-1&0&0&0&0&0 \\
  0&0&0&-1&0&0&0&0 \\
  1&0&0&0&0&0&0&0 \\
  0&1&0&0&0&0&0&0 \\
  0&0&0&0&0&0&1&0 \\
  0&0&0&0&0&0&0&1 \\
  0&0&0&0&-1&0&0&0 \\
  0&0&0&0&0&-1&0&0 \\
\end{pmatrix},$

$A_4=
\begin{pmatrix}
  0&0&0&-1&0&0&0&0 \\
  0&0&1&0&0&0&0&0 \\
  0&-1&0&0&0&0&0&0 \\
  1&0&0&0&0&0&0&0 \\
  0&0&0&0&0&0&0&1 \\
  0&0&0&0&0&0&-1&0 \\
  0&0&0&0&0&1&0&0 \\
  0&0&0&0&-1&0&0&0 \\
\end{pmatrix},$

$A_5=
\begin{pmatrix}
  0&0&0&0&-1&0&0&0 \\
  0&0&0&0&0&-1&0&0 \\
  0&0&0&0&0&0&-1&0 \\
  0&0&0&0&0&0&0&-1 \\
  1&0&0&0&0&0&0&0 \\
  0&1&0&0&0&0&0&0 \\
  0&0&1&0&0&0&0&0 \\
  0&0&0&1&0&0&0&0 \\
\end{pmatrix},$

$A_6=
\begin{pmatrix}
  0&0&0&0&0&-1&0&0 \\
  0&0&0&0&1&0&0&0 \\
  0&0&0&0&0&0&0&-1 \\
  0&0&0&0&0&0&1&0 \\
  0&-1&0&0&0&0&0&0 \\
  1&0&0&0&0&0&0&0 \\
  0&0&0&-1&0&0&0&0 \\
  0&0&1&0&0&0&0&0 \\
\end{pmatrix},$

$A_7=
\begin{pmatrix}
  0&0&0&0&0&0&-1&0 \\
  0&0&0&0&0&0&0&1 \\
  0&0&0&0&1&0&0&0 \\
  0&0&0&0&0&-1&0&0 \\
  0&0&-1&0&0&0&0&0 \\
  0&0&0&1&0&0&0&0 \\
  1&0&0&0&0&0&0&0 \\
  0&-1&0&0&0&0&0&0 \\
\end{pmatrix},$

$A_8=
\begin{pmatrix}
  0&0&0&0&0&0&0&-1 \\
  0&0&0&0&0&0&-1&0 \\
  0&0&0&0&0&1&0&0 \\
  0&0&0&0&1&0&0&0 \\
  0&0&0&-1&0&0&0&0 \\
  0&0&-1&0&0&0&0&0 \\
  0&1&0&0&0&0&0&0 \\
  1&0&0&0&0&0&0&0 \\
\end{pmatrix}.\\
$\\


\begin{thebibliography}{24}
\expandafter\ifx\csname natexlab\endcsname\relax\def\natexlab#1{#1}\fi
\expandafter\ifx\csname bibnamefont\endcsname\relax
  \def\bibnamefont#1{#1}\fi
\expandafter\ifx\csname bibfnamefont\endcsname\relax
  \def\bibfnamefont#1{#1}\fi
\expandafter\ifx\csname citenamefont\endcsname\relax
  \def\citenamefont#1{#1}\fi
\expandafter\ifx\csname url\endcsname\relax
  \def\url#1{\texttt{#1}}\fi
\expandafter\ifx\csname urlprefix\endcsname\relax\def\urlprefix{URL }\fi
\providecommand{\bibinfo}[2]{#2}
\providecommand{\eprint}[2][]{\url{#2}}

\bibitem[{\citenamefont{Gisin et~al.}(2002)\citenamefont{Gisin, Ribordy,
  Tittel, and Zbinden}}]{gisin:rmp}
\bibinfo{author}{\bibfnamefont{N.}~\bibnamefont{Gisin}},
  \bibinfo{author}{\bibfnamefont{G.}~\bibnamefont{Ribordy}},
  \bibinfo{author}{\bibfnamefont{W.}~\bibnamefont{Tittel}}, \bibnamefont{and}
  \bibinfo{author}{\bibfnamefont{H.}~\bibnamefont{Zbinden}},
  \bibinfo{journal}{Rev. Mod. Phys.} \textbf{\bibinfo{volume}{74}},
  \bibinfo{pages}{145} (\bibinfo{year}{2002}).

\bibitem[{\citenamefont{Grosshans et~al.}(2003)\citenamefont{Grosshans, Assche,
  Wenger, Brouri, Cerf, and Grangier}}]{grosshans:nature}
\bibinfo{author}{\bibfnamefont{F.}~\bibnamefont{Grosshans}},
  \bibinfo{author}{\bibfnamefont{G.~V.} \bibnamefont{Assche}},
  \bibinfo{author}{\bibfnamefont{J.}~\bibnamefont{Wenger}},
  \bibinfo{author}{\bibfnamefont{R.}~\bibnamefont{Brouri}},
  \bibinfo{author}{\bibfnamefont{N.~J.} \bibnamefont{Cerf}}, \bibnamefont{and}
  \bibinfo{author}{\bibfnamefont{P.}~\bibnamefont{Grangier}},
  \bibinfo{journal}{Nature} \textbf{\bibinfo{volume}{421}},
  \bibinfo{pages}{238} (\bibinfo{year}{2003}).

\bibitem[{\citenamefont{Bennett et~al.}(1984)\citenamefont{Bennett, Brassard
  et~al.}}]{bb84}
\bibinfo{author}{\bibfnamefont{C.}~\bibnamefont{Bennett}},
  \bibinfo{author}{\bibfnamefont{G.}~\bibnamefont{Brassard}},
  \bibnamefont{et~al.}, \bibinfo{journal}{Proceedings of IEEE International
  Conference on Computers, Systems, and Signal Processing}
  \textbf{\bibinfo{volume}{175}} (\bibinfo{year}{1984}).

\bibitem[{\citenamefont{{Van~Assche} et~al.}(2004)\citenamefont{{Van~Assche},
  Cardinal, and Cerf}}]{VanAssche2004}
\bibinfo{author}{\bibfnamefont{G.}~\bibnamefont{{Van~Assche}}},
  \bibinfo{author}{\bibfnamefont{J.}~\bibnamefont{Cardinal}}, \bibnamefont{and}
  \bibinfo{author}{\bibfnamefont{N.~J.} \bibnamefont{Cerf}},
  \bibinfo{journal}{IEEE Transactions on Information Theory}
  \textbf{\bibinfo{volume}{50}}, \bibinfo{pages}{394} (\bibinfo{year}{2004}).

\bibitem[{\citenamefont{Bloch et~al.}(2006)\citenamefont{Bloch, Thangaraj,
  McLaughlin, and Merolla}}]{Bloch2006}
\bibinfo{author}{\bibfnamefont{M.}~\bibnamefont{Bloch}},
  \bibinfo{author}{\bibfnamefont{A.}~\bibnamefont{Thangaraj}},
  \bibinfo{author}{\bibfnamefont{S.~W.} \bibnamefont{McLaughlin}},
  \bibnamefont{and} \bibinfo{author}{\bibfnamefont{J.-M.}
  \bibnamefont{Merolla}}, in \emph{\bibinfo{booktitle}{Proc. I{EEE}
  {I}nformation {T}heory {W}orkshop}} (\bibinfo{address}{Punta del Este,
  Uruguay}, \bibinfo{year}{2006}), \bibinfo{note}{arXiv:cs.IT/0509041}.

\bibitem[{\citenamefont{Ralph}(2000)}]{ralph}
\bibinfo{author}{\bibfnamefont{T.~C.} \bibnamefont{Ralph}},
  \bibinfo{journal}{Phys. Rev. A} \textbf{\bibinfo{volume}{61}},
  \bibinfo{pages}{010303(R)} (\bibinfo{year}{2000}).

\bibitem[{\citenamefont{Silberhorn et~al.}(2002)\citenamefont{Silberhorn,
  Ralph, L\"{u}tkenhaus, and Leuchs}}]{silberhorn2002}
\bibinfo{author}{\bibfnamefont{C.}~\bibnamefont{Silberhorn}},
  \bibinfo{author}{\bibfnamefont{T.}~\bibnamefont{Ralph}},
  \bibinfo{author}{\bibfnamefont{N.}~\bibnamefont{L\"{u}tkenhaus}},
  \bibnamefont{and} \bibinfo{author}{\bibfnamefont{G.}~\bibnamefont{Leuchs}},
  \bibinfo{journal}{Phys. Rev. Lett.} \textbf{\bibinfo{volume}{89}},
  \bibinfo{pages}{167901} (\bibinfo{year}{2002}).

\bibitem[{\citenamefont{Weedbrook et~al.}(2004)\citenamefont{Weedbrook, Lance,
  Bowen, Symul, Ralph, and Lam}}]{R04}
\bibinfo{author}{\bibfnamefont{C.}~\bibnamefont{Weedbrook}},
  \bibinfo{author}{\bibfnamefont{A.}~\bibnamefont{Lance}},
  \bibinfo{author}{\bibfnamefont{W.}~\bibnamefont{Bowen}},
  \bibinfo{author}{\bibfnamefont{T.}~\bibnamefont{Symul}},
  \bibinfo{author}{\bibfnamefont{T.}~\bibnamefont{Ralph}}, \bibnamefont{and}
  \bibinfo{author}{\bibfnamefont{P.}~\bibnamefont{Lam}},
  \bibinfo{journal}{Phys. Rev. Lett.} \textbf{\bibinfo{volume}{93}},
  \bibinfo{pages}{170504} (\bibinfo{year}{2004}).

\bibitem[{\citenamefont{Namiki and Hirano}(2004)}]{namiki-2004}
\bibinfo{author}{\bibfnamefont{R.}~\bibnamefont{Namiki}} \bibnamefont{and}
  \bibinfo{author}{\bibfnamefont{T.}~\bibnamefont{Hirano}},
  \bibinfo{journal}{Phys. Rev. Lett.} \textbf{\bibinfo{volume}{92}},
  \bibinfo{pages}{117901} (\bibinfo{year}{2004}).

\bibitem[{\citenamefont{Weedbrook et~al.}(2005)\citenamefont{Weedbrook, Lance,
  Bowen, Symul, Ralph, and Lam}}]{2005no-}
\bibinfo{author}{\bibfnamefont{C.}~\bibnamefont{Weedbrook}},
  \bibinfo{author}{\bibfnamefont{A.}~\bibnamefont{Lance}},
  \bibinfo{author}{\bibfnamefont{W.}~\bibnamefont{Bowen}},
  \bibinfo{author}{\bibfnamefont{T.}~\bibnamefont{Symul}},
  \bibinfo{author}{\bibfnamefont{T.}~\bibnamefont{Ralph}}, \bibnamefont{and}
  \bibinfo{author}{\bibfnamefont{P.}~\bibnamefont{Lam}},
  \bibinfo{journal}{Phys. Rev. Lett.} \textbf{\bibinfo{volume}{95}},
  \bibinfo{pages}{180503} (\bibinfo{year}{2005}).

\bibitem[{\citenamefont{Heid and L\"utkenhaus}(2007)}]{Heid2007}
\bibinfo{author}{\bibfnamefont{M.}~\bibnamefont{Heid}} \bibnamefont{and}
  \bibinfo{author}{\bibfnamefont{N.}~\bibnamefont{L\"utkenhaus}},
  \bibinfo{journal}{Phys. Rev. A} \textbf{\bibinfo{volume}{76}},
  \bibinfo{pages}{022313} (\bibinfo{year}{2007}),
  \eprint{arXiv:quant-ph/0608015}.

\bibitem[{\citenamefont{Garcia-Patron and Cerf}(2006)}]{garcia-patron:prl}
\bibinfo{author}{\bibfnamefont{R.}~\bibnamefont{Garcia-Patron}}
  \bibnamefont{and} \bibinfo{author}{\bibfnamefont{N.~J.} \bibnamefont{Cerf}},
  \bibinfo{journal}{Phys. Rev. Lett.} \textbf{\bibinfo{volume}{97}},
  \bibinfo{pages}{190503} (\bibinfo{year}{2006}).

\bibitem[{\citenamefont{Navascu\'es et~al.}(2006)\citenamefont{Navascu\'es,
  Grosshans, and Ac\'in}}]{navascues:prl06}
\bibinfo{author}{\bibfnamefont{M.}~\bibnamefont{Navascu\'es}},
  \bibinfo{author}{\bibfnamefont{F.}~\bibnamefont{Grosshans}},
  \bibnamefont{and} \bibinfo{author}{\bibfnamefont{A.}~\bibnamefont{Ac\'in}},
  \bibinfo{journal}{Phys. Rev. Lett.} \textbf{\bibinfo{volume}{97}},
  \bibinfo{pages}{190502} (\bibinfo{year}{2006}).

\bibitem[{\citenamefont{Cover and Thomas}(1991)}]{cover}
\bibinfo{author}{\bibfnamefont{T.~M.} \bibnamefont{Cover}} \bibnamefont{and}
  \bibinfo{author}{\bibfnamefont{J.~A.} \bibnamefont{Thomas}},
  \emph{\bibinfo{title}{Elements of Information Theory}}
  (\bibinfo{publisher}{Wiley-Interscience}, \bibinfo{year}{1991}).

\bibitem[{\citenamefont{Nielsen and Chuang}(2000)}]{Nielsen}
\bibinfo{author}{\bibfnamefont{M.~A.} \bibnamefont{Nielsen}} \bibnamefont{and}
  \bibinfo{author}{\bibfnamefont{I.~L.} \bibnamefont{Chuang}},
  \emph{\bibinfo{title}{Quantum Information and Quantum Computation}}
  (\bibinfo{publisher}{Cambridge University Press}, \bibinfo{year}{2000}).

\bibitem[{\citenamefont{Renner}(2005)}]{R05}
\bibinfo{author}{\bibfnamefont{R.}~\bibnamefont{Renner}}, Ph.D. thesis,
  \bibinfo{school}{ETH Zurich} (\bibinfo{year}{2005}).

\bibitem[{\citenamefont{Renner}(2007)}]{renner2007}
\bibinfo{author}{\bibfnamefont{R.}~\bibnamefont{Renner}},
  \bibinfo{journal}{Nature Physics} \textbf{\bibinfo{volume}{3}},
  \bibinfo{pages}{645} (\bibinfo{year}{2007}).

\bibitem[{\citenamefont{Shannon}(1948)}]{Sha48}
\bibinfo{author}{\bibfnamefont{C.}~\bibnamefont{Shannon}},
  \bibinfo{journal}{Bell System Technical Journal}
  \textbf{\bibinfo{volume}{27}}, \bibinfo{pages}{379} (\bibinfo{year}{1948}).

\bibitem[{\citenamefont{Berrou et~al.}(1993)\citenamefont{Berrou, Glavieux, and
  Thitimajshima}}]{berrou}
\bibinfo{author}{\bibfnamefont{C.}~\bibnamefont{Berrou}},
  \bibinfo{author}{\bibfnamefont{A.}~\bibnamefont{Glavieux}}, \bibnamefont{and}
  \bibinfo{author}{\bibfnamefont{P.}~\bibnamefont{Thitimajshima}},
  \bibinfo{journal}{Communications, 1993. ICC 93. Geneva. Technical Program,
  Conference Record, IEEE International Conference on}
  \textbf{\bibinfo{volume}{2}} (\bibinfo{year}{1993}).

\bibitem[{\citenamefont{Richardson et~al.}(2001)\citenamefont{Richardson,
  Shokrollahi, and Urbanke}}]{Richardson2001a}
\bibinfo{author}{\bibfnamefont{T.~J.} \bibnamefont{Richardson}},
  \bibinfo{author}{\bibfnamefont{M.~A.} \bibnamefont{Shokrollahi}},
  \bibnamefont{and} \bibinfo{author}{\bibfnamefont{R.~L.}
  \bibnamefont{Urbanke}}, \bibinfo{journal}{IEEE Transactions on Information
  Theory} \textbf{\bibinfo{volume}{47}}, \bibinfo{pages}{619}
  (\bibinfo{year}{2001}).

\bibitem[{\citenamefont{Wyner}(1975)}]{Wyner}
\bibinfo{author}{\bibfnamefont{A.~D.} \bibnamefont{Wyner}},
  \bibinfo{journal}{Bell System Technical Journal}
  \textbf{\bibinfo{volume}{54}}, \bibinfo{pages}{1355} (\bibinfo{year}{1975}).

\bibitem[{\citenamefont{Lodewyck et~al.}(2007)\citenamefont{Lodewyck, Bloch,
  Garc\'ia-Patr\'on, Fossier, Karpov, Diamanti, Debuisschert, Cerf,
  Tualle-Brouri, McLaughlin et~al.}}]{lodewyck:pra07}
\bibinfo{author}{\bibfnamefont{J.}~\bibnamefont{Lodewyck}},
  \bibinfo{author}{\bibfnamefont{M.}~\bibnamefont{Bloch}},
  \bibinfo{author}{\bibfnamefont{R.}~\bibnamefont{Garc\'ia-Patr\'on}},
  \bibinfo{author}{\bibfnamefont{S.}~\bibnamefont{Fossier}},
  \bibinfo{author}{\bibfnamefont{E.}~\bibnamefont{Karpov}},
  \bibinfo{author}{\bibfnamefont{E.}~\bibnamefont{Diamanti}},
  \bibinfo{author}{\bibfnamefont{T.}~\bibnamefont{Debuisschert}},
  \bibinfo{author}{\bibfnamefont{N.~J.} \bibnamefont{Cerf}},
  \bibinfo{author}{\bibfnamefont{R.}~\bibnamefont{Tualle-Brouri}},
  \bibinfo{author}{\bibfnamefont{S.~W.} \bibnamefont{McLaughlin}},
  \bibnamefont{et~al.}, \bibinfo{journal}{Phys. Rev. A}
  \textbf{\bibinfo{volume}{76}}, \bibinfo{pages}{042305}
  (\bibinfo{year}{2007}).

\bibitem[{\citenamefont{Adams}(1962)}]{adams}
\bibinfo{author}{\bibfnamefont{J.}~\bibnamefont{Adams}},
  \bibinfo{journal}{Annals of Math.} \textbf{\bibinfo{volume}{75}},
  \bibinfo{pages}{603} (\bibinfo{year}{1962}).

\bibitem[{\citenamefont{http://ltchwww.epfl.ch/research/ldpcopt}()}]{epfl}
\bibinfo{author}{\bibnamefont{http://ltchwww.epfl.ch/research/ldpcopt}}.

\end{thebibliography}
\end{document}